\documentclass[letterpaper]{jpconf}
\usepackage[latin1]{inputenc}
\usepackage{amsmath,amssymb}
\usepackage{graphicx} 

\newcommand{\gras}[1]{\boldsymbol{#1}}

\begin{document}

\title{\bf Computing Heavy Elements}

\author{N. Schunck$^{1}$, A. Baran$^{2,6}$, M. Kortelainen$^{2,3}$, 
        J. McDonnell$^{2,3}$, J. Mor\'{e}$^{5}$, W. Nazarewicz$^{2,3,4}$, 
        J. Pei$^{2,3}$, J. Sarich$^{5}$, J. Sheikh$^{2,3}$, A. Staszczak$^{2,6}$, 
        M. Stoitsov$^{2,3}$, and S.M. Wild$^{5}$}

\address{$^{1}$ Physics Division, Lawrence Livermore National Laboratory 
         Livermore, CA 94551, USA}
\address{$^{2}$ Department of Physics and Astronomy, University of Tennessee, 
         Knoxville, TN 37996, USA}
\address{$^{3}$ Physics Division, Oak Ridge National Laboratory, P.O. Box 2008, 
         Oak Ridge, TN 37831, USA}
\address{$^{4}$ Institute of Theoretical Physics, Warsaw University, ul. Ho\.{z}a 
         69, PL-00681, Warsaw, Poland}
\address{$^{5}$ Mathematics and Computer Science Division, Argonne National 
         Laboratory, Argonne, IL 60439, USA}
\address{$^{6}$ Department of Theoretical Physics, Maria Curie-Sk{\l}odowska 
         University, pl. M. Curie-Sk{\l}odowskiej 1, 20-031 Lublin, Poland}

\ead{schunck1@llnl.gov}

\begin{abstract}
Reliable calculations of the structure of heavy elements are crucial to 
address fundamental science questions such as the origin of the elements 
in the universe. Applications relevant for energy production, medicine, 
or national security also rely on theoretical predictions of basic 
properties of atomic nuclei. Heavy elements are best described within the 
nuclear density functional theory (DFT) and its various extensions. While 
relatively mature, DFT has never been implemented in its full power, as 
it relies on a very large number ($\sim 10^{9}-10^{12}$) of expensive 
calculations ($\sim$ day). The advent of leadership-class computers, as 
well as dedicated large-scale collaborative efforts such as the SciDAC 2 
UNEDF project, have dramatically changed the field. This article gives 
an overview of the various computational challenges related to the nuclear 
DFT, as well as some of the recent achievements.
\end{abstract}


\section{Introduction}

Nuclear physics pervades a number of scientific disciplines as well as 
societal applications. Understanding the production of the elements in 
stellar interiors is key to reproduce the observed isotopic abundances in 
the universe. It necessitates an accurate knowledge of the structure of 
radioactive nuclides that are so short-lived that they will remain beyond 
reach of experimental facilities for many years to come, if not for ever. In 
the long term, the safety and viability of nuclear energy production sources 
will be enhanced by acquiring a precise understanding of the complex 
mechanisms of fission and fusion. The latter are also critical for stockpile 
stewardship, which has broad implications for national security. 

At the heart of these important issues lies the elusive structure of the 
atomic nucleus. From a physics standpoint, it is a quantum many-body 
problem facing three major theoretical challenges: (1) the interaction that 
binds neutrons and protons together in the nucleus is in principle derived 
from quantum chromodynamics (QCD), but this derivation has not been firmly 
established yet; (2) the interaction between nucleons inside the nucleus 
is very different from the one between isolated nucleons in the vacuum 
(in-medium interactions); (3) the number of constituents in the nucleus 
($\sim$1--300) almost always forbids both exact analytical solutions, except 
for the lightest systems, as well as the use of statistical methods 
applicable to systems with a very large number of particles. In spite of 
these formidable difficulties, there has been significant progress over 
the past 50 years to address all these issues. 

While there exist many excellent models for light nuclei, heavy elements can 
be described only by what is variously known as the nuclear self-consistent 
mean-field theory or, more recently, density functional theory (DFT) 
\cite{[Ben03]}. Since its inception in the 1950ies, DFT has reached a 
satisfactory level of maturity. Until now, however, computational limitations 
did not allow to implement the theory as originally designed, resulting in 
uncontrolled systematic errors, poor precision, and dubious reliability in 
regions of exotic nuclei where experimental information is scarce or 
nonexistent. The fast development of leadership-class computers has for 
the first time lifted many of these limitations, and the solution to 
long-standing problems seems now possible in the short term. After briefly 
introducing the underlying theoretical background, this article discusses 
some of the computational challenges and methods used in nuclear DFT, 
highlights some of the recent achievements, and discusses current open 
problems. 


\section{Theoretical Models of Heavy Nuclei}

The central hypothesis of nuclear DFT is that the $A$ nucleons (protons and 
neutrons) inside the nucleus can be treated as independent quasi-particles 
moving in an average nuclear potential well. The theory can be entirely 
formulated by introducing the so-called one-body density matrix $\rho(x,x')$ 
and pairing tensor $\kappa(x,x')$, where $x \equiv (\gras{r},\sigma)$ 
includes spatial as well as spin coordinates, $\sigma=\pm 1/2$. Requiring 
that the total energy $E$ of the nucleus is minimal under a variation of both 
$\rho$ and $\kappa$ leads to the so-called Hartree-Fock-Bogoliubov (HFB) 
equations:
\begin{equation}
\int dx'
\left(
\begin{array}{cc}
h[\rho(x,x')] - \lambda   & \Delta[\kappa(x,x')] \\
-\Delta^{*}[\kappa(x,x')] & -h^{*}[\rho(x,x')] + \lambda
\end{array}
\right)
\left(
\begin{array}{c}
U_{\mu}(x') \\
V_{\mu}(x')
\end{array}
\right)
=
E_{\mu}
\left(
\begin{array}{c}
U_{\mu}(x) \\
V_{\mu}(x)
\end{array}
\right), \ \ \mu=1,\dots,+\infty,
\label{eq:hfb}
\end{equation}
with $\lambda$ a Lagrange parameter that must be introduced to conserve 
particle number, and
\begin{equation}
\begin{array}{l}
  \rho(x,x') = \displaystyle\sum_{\mu} V_{\mu}^{*}(x)V_{\mu}(x'), \medskip\\
\kappa(x,x') = \displaystyle\sum_{\mu} V_{\mu}^{*}(x)U_{\mu}(x').
\end{array}
\end{equation}
In Eq. (\ref{eq:hfb}), $h[\rho]$ is a Hermitian operator (mean-field), which 
is a functional of the density matrix, and $\Delta[\kappa]$ is an 
antisymmetric operator (pairing field), which is a functional of the pairing 
tensor. The explicit dependence of the HFB matrix on the eigenfunctions 
$(U_{\mu},V_{\mu})$ via the density matrix and pairing tensor, or 
self-consistency, makes the eigenvalue problem highly nonlinear.

In its most general form, the mean-field operator reads:
\begin{equation}
h = 
-\frac{\hbar^{2}}{2m} \gras{\nabla}^{2} + \Gamma(x,x')
\end{equation}
with $\hbar$ the Planck constant, $m$ the mass of a nucleon, $\gras{\nabla}$ 
the gradient operator, and $\Gamma$ the so-called Hartree-Fock potential. 
In phenomenological mean-field models, $\Gamma$ is in fact parametrized by 
some suitable mathematical function and does not depend on the density 
matrix. In the traditional version of the self-consistent mean-field theory, 
$\Gamma$ is instead computed from a {\it local} two-body interaction, or 
effective pseudo-potential, $V(x,x')$, which depends on the spatial and spin 
coordinates $x$ and $x'$ of two nucleons and takes the general form
\begin{equation}
\Gamma(x,x') = \delta(x-x')\int dx_{1} V(x,x_{1})\rho(x_{1},x_{1}) - 
V(x,x')\rho(x,x').
\label{eq:gammaR}
\end{equation}
Standard two-body interactions have either zero-range, that is, 
$V(x,x') \sim \delta(\gras{r} - \gras{r}')$ (Skyrme forces), or 
finite-range, $V(x,x') \sim  V(\gras{r} - \gras{r}')$ (Gogny forces), 
which may further simplify the general expression (\ref{eq:gammaR}). 
They all contain a density-dependent term that is necessary to reproduce the 
saturation of nuclear matter but that makes them ill-behaved for extensions 
of the theory dealing with large amplitude collective motion \cite{[Dob07d],
[Dug10]}. Recent formulations of nuclear DFT do not consider explicitly 
effective interactions and instead parametrize $\Gamma(x,x')$ directly as a functional of the density matrix $\rho(x,x')$, or alternatively the 
local density $\rho(x)$ and its spatial derivatives.
		
The success of the nuclear mean-field theory relies on the mechanism of 
spontaneous symmetry breaking, whereby the solutions of the HFB equations 
(\ref{eq:hfb}) may break some of the symmetries of the underlying effective 
interaction $V(x,x')$. This mechanism can be viewed as a way to introduce 
correlations in what is otherwise an independent particle model. A simple 
example is the breaking of rotational invariance of $V(x,x')$, which 
implies that the density matrix and pairing tensor can have nonisotropic 
spatial distributions: in the mean-field theory, nuclei can be deformed, and 
the energy of the nucleus therefore depends on the deformation. However, this 
dependence is not known beforehand. In practice, one must introduce 
constraint operators $\hat{Q}_{lm}$ to probe the deformation energy surface. 
The problem is complicated by the fact that the expectation value $\langle 
Q_{lm}\rangle$ of the (local) constraint operator $\hat{Q}_{lm}$ is itself a 
functional of the density matrix, 
\begin{equation}
\langle Q_{lm}\rangle = \int dx \hat{Q}_{lm}(x)\rho(x,x).
\label{eq:constraints}
\end{equation}
The HFB equations (\ref{eq:hfb}) with the constraints (\ref{eq:constraints}) 
represent a system of coupled, nonlinear, integro-differential equations 
and are the cornerstone of the description of heavy elements in a microscopic 
framework. In the following, we discuss the various methods used to solve 
these equations, as well as the related mathematical and computational challenges.


\section{Mathematical and Computational Challenges}

From a computational perspective, nuclear DFT has two facets: (i) the HFB 
solver itself and (ii) the management of a large number of quasi-independent, 
time-consuming, load-imbalanced tasks. We discuss below each of these 
aspects.


\subsection{HFB Solver}

{\sc Solving the HFB Equations - } There exist essentially two classes of 
methods to solve the HFB equations. In the coordinate representation, the 
equations (\ref{eq:hfb}) are solved directly by numerical integration for 
each eigenfunction $\mu$. Boundary conditions for the wave-functions are 
imposed on the domain of integration. While precise, the feasibility 
and usability of this approach are highly dependent on the symmetries of the 
wave functions: in spherical symmetry, the eigenfunctions are separable 
$(U_{\mu}(\gras{r}),V_{\mu}(\gras{r})) = (u_{\mu}(r),v_{\mu}(r))Y_{lm}(\theta,
\varphi)$. The HFB equations depend only on the radial coordinate $r$, and 
numerical implementations can be very fast (typically less than 1 minute per 
HFB calculation) \cite{[Ben05]}. In cylindrical symmetry, the wave-functions 
depend explicitly on the two variables $z$ and $\rho$, and special separation 
techniques (B-splines or similar) must be employed. Codes with built-in 
parallel capabilities achieve good convergence for a few dozens of cores/HFB 
calculation in a few hours \cite{[Pei08]}. Full 3D solvers in coordinate space 
are in development: they will probably require a large number of cores 
($>$1,000) to achieve convergence in less than a day. At this scale, the 
benefits of using DFT (reformulate the problem to replace the $3A$ coordinates 
of the nucleons by the only three coordinates of the density matrix) are greatly 
reduced though. 

\begin{figure}[h]
\begin{center}
\begin{minipage}[t]{0.389\textwidth}
\begin{center}
  \includegraphics[width=0.95\linewidth]{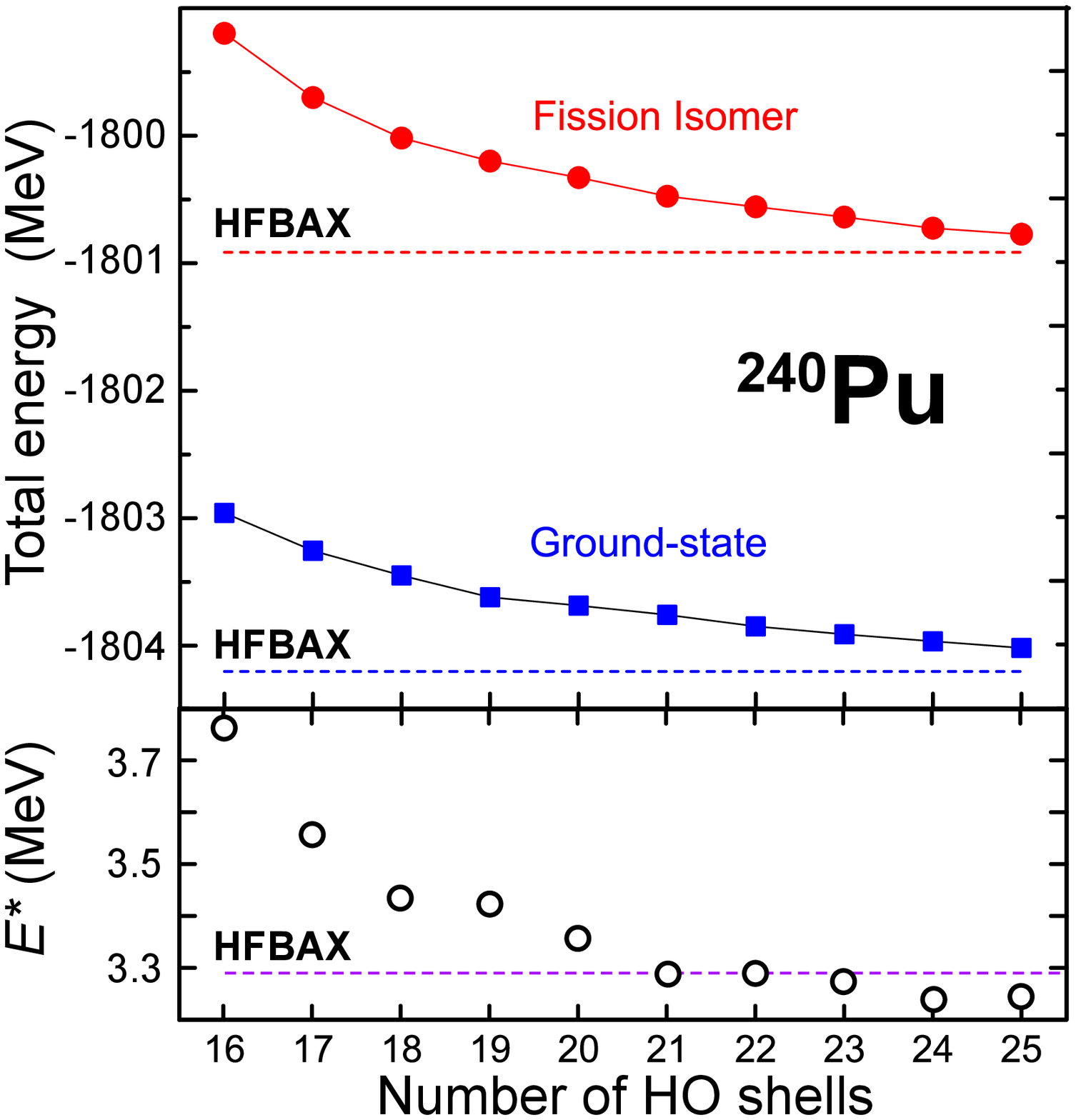}
  \caption{(color online) Convergence of the energy for various deformed 
           states in $^{240}$Pu as function of the number of HO shells. The 
           dashed line gives the result of \cite{[Pei08]}.}
  \label{fig:numerics}
\end{center}
\end{minipage}\hspace{0.03\textwidth}%
\begin{minipage}[t]{0.57\textwidth}
\begin{center}
  \includegraphics[width=1.\linewidth]{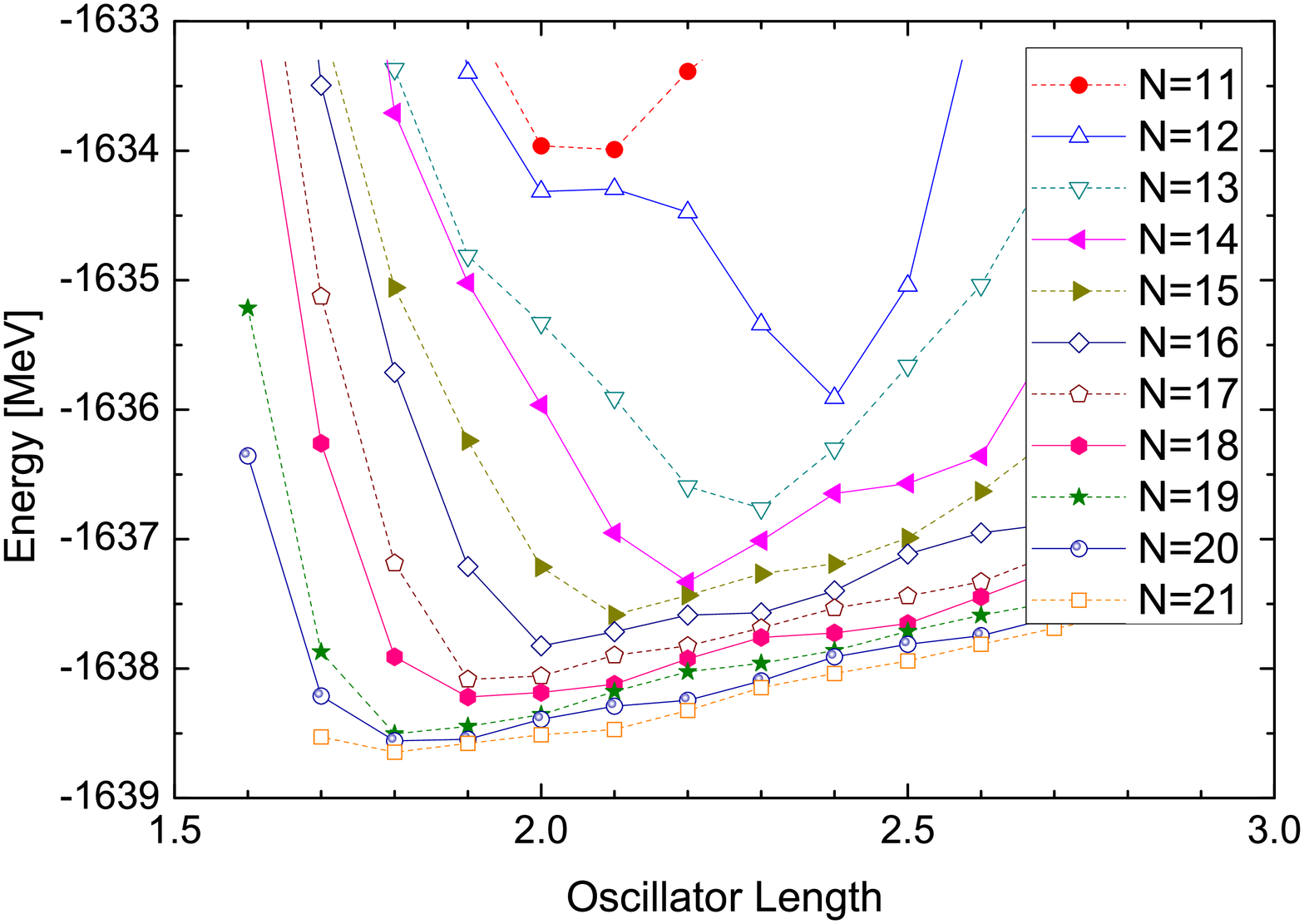}
  \caption{(color online) Convergence of the energy for the ground-state 
           of $^{208}$Pb as function of the HO oscillator length 
           $b = \sqrt{m\omega_{0}/\hbar}$ for various number of 
           oscillator shells.}
  \label{fig:numerics-1}
\end{center}
\end{minipage}
\end{center}
\end{figure}

The alternative approach consists of introducing a basis of the Hilbert 
space $\mathcal{L}_{2}$ of square-integrable functions and computing the 
matrix elements of all relevant operators in that basis. The Harmonic 
Oscillator (HO) basis proves the most adapted to nuclear structure 
applications as it involves very localized basis functions. By comparison, 
molecular physics applications often employ the plane wave basis. The 
choice of the coordinate system is dictated by the symmetries that one wants 
to impose, or relax, on the system. For example, fission studies typically 
require many spatial symmetries to be broken, and the Cartesian HO basis 
becomes the tool of choice. The HO basis is by definition an infinite countable 
basis of the one-particle Hilbert space: numerical implementations require the 
truncation of the expansion up to a maximum number of oscillator shells $N$. 
This introduces systematic truncation errors, as well as an artificial dependence 
on the frequency $\omega_{0}$ of the HO. When studying very deformed nuclei, 
it is recommended to choose an anisotropic 2D HO, with $\omega_{\perp} \neq 
\omega_{\parallel}$. The final truncation error then depends also on the 
deformation of the basis, that is, the ratio $q = \omega_{\parallel} / 
\omega_{\perp}$. For any practical calculation, the final HFB energy $E$ 
should therefore be written $E(N, \omega_{0}, q)$. High-precision 
calculations require systematic and costly studies of this model space 
dependence; see Figs. \ref{fig:numerics}--\ref{fig:numerics-1} \cite{[Nik10]}. 
Recent attempts have therefore been made to apply multiresolution methods 
based on wavelet expansions to nuclear DFT, in order to combine the 
versatility of basis expansion with arbitrary precision results 
\cite{[Fan09]}.

{\sc Dense Linear Algebra - } When expressed in a single-particle basis, the 
density $\rho(x,x')$ becomes 
an actual matrix $\rho_{ij}$ defined as
\begin{equation}
\rho_{ij} = \iint dxdx' \phi_{j}^{*}(x')\rho(x,x')\phi_{i}(x).
\end{equation}
A general two-body interaction becomes a rank-4 tensor $v_{abcd}$. The 
HF potential is obtained by taking a tensor contraction:
\begin{equation}
\Gamma_{ac} = \sum_{bd} v_{abcd}\rho_{db},
\label{eq:gammaHO}
\end{equation}
where the summation for each index extends over the size of the basis. Such 
tensor contractions must be performed at every iteration and represent an 
important bottleneck in the calculation. Indeed, for a heavy nucleus, 
the size of the HO basis for a precise calculation contains typically $N>20$ 
shells. In a Cartesian basis, this implies that each index $a, b, c, d$ is 
in fact a set of three numbers, $a \rightarrow \gras{n}_{a} \equiv (n_{x,a},n_{y,a}, 
n_{z,a})$ with $n_{x,a} + n_{y,a} + n_{z,a} = N$. A naive implementation of Eq. 
(\ref{eq:gammaHO}) would therefore require a 12-nested loop to compute the 
entire matrix $\Gamma_{ac}$. Most problematic, in double-precision arithmetic 
the size of the complex tensor $v_{abcd}$ would be on the order of 80 TB 
after taking into account the anti-symmetry properties of $v_{abcd}$. Current 
implementations therefore do not store the matrix $v_{abcd}$ but compute it 
on the fly, adding to the computational overhead, and store the matrix 
$\Gamma_{ac}$ instead ($\sim$ 50 MB storage).

An alternative technique to compute $\Gamma_{ac}$, which avoids handling 
directly the tensor $v_{abcd}$ and is particularly efficient when the 
functional depends only on the local density matrix, relies on the fact that 
the functional dependence of $\Gamma$ on the density matrix $\rho$ is known. 
The HF potential for a local functional is simplified: $\Gamma[\rho(x,x')] 
\rightarrow \Gamma[\rho(x)]$. The matrix $\Gamma_{ac}$ can then be computed 
by only one 3D integral:
\begin{equation}
\Gamma_{ac} = \int dx \phi_{j}^{*}(x)\Gamma[\rho(x)]\phi_{i}(x).
\end{equation}
Such integrations can be performed exactly by quadrature formulas. The only 
time-consuming part of this method is to obtain an expression of $\rho(x)$ on 
the quadrature mesh, that is, $\rho(x_{k_{x}}, y_{k_{y}}, z_{k_{z}})$. One must compute
\begin{equation}
\rho(x_{k_{x}}, y_{k_{y}}, z_{k_{z}}) = 
\sum_{\gras{m}\gras{n}}
\sum_{\mu} V_{\gras{m}\mu}^{*}V_{\gras{n}\mu}
\phi_{\gras{n}}^{*}(x_{k_{x}}, y_{k_{y}}, z_{k_{z}})
\phi_{\gras{m}}(x_{k_{x}}, y_{k_{y}}, z_{k_{z}}).
\end{equation}
This dense linear algebra requires at first sight 10-nested loops to 
build the entire representation of $\rho$ on the integration grid. 
Various numerical tricks can be used to reduce this number \cite{[Dob09d]}. 
		

\begin{figure}[h]
\begin{center}
\begin{minipage}[t]{0.48\textwidth}
\begin{center}
  \includegraphics[width=1.0\linewidth]{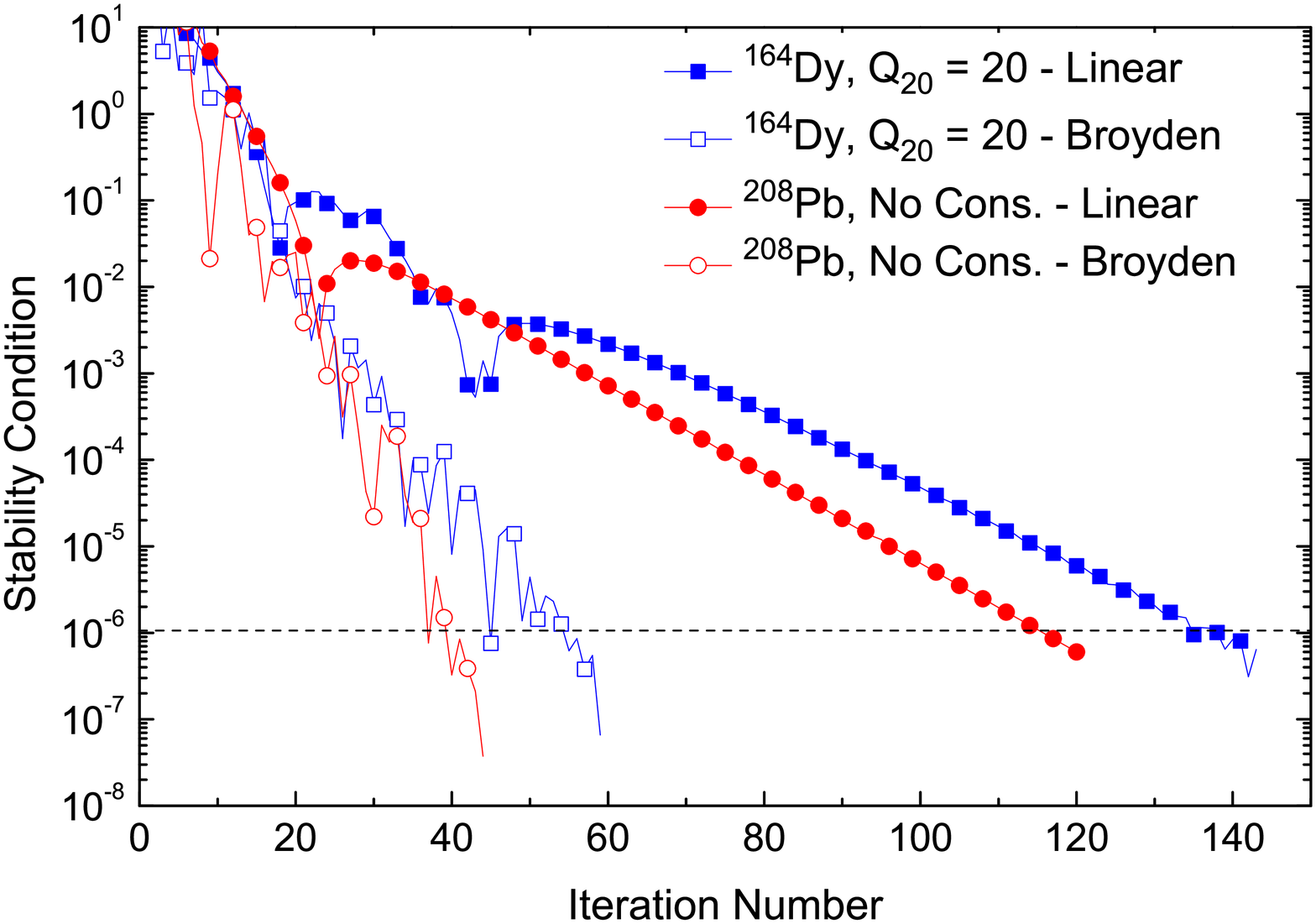}
  \caption{(color online) Convergence of the energy of two different nuclei with 
           and without constraint on the quadrupole moment using standard 
           linear mixing and the modified Broyden method \cite{[Bar08]}.}
  \label{fig:broyden}
\end{center}
\end{minipage}\hspace{0.03\textwidth}%
\begin{minipage}[t]{0.48\textwidth}
\begin{center}
  \includegraphics[width=0.93\linewidth]{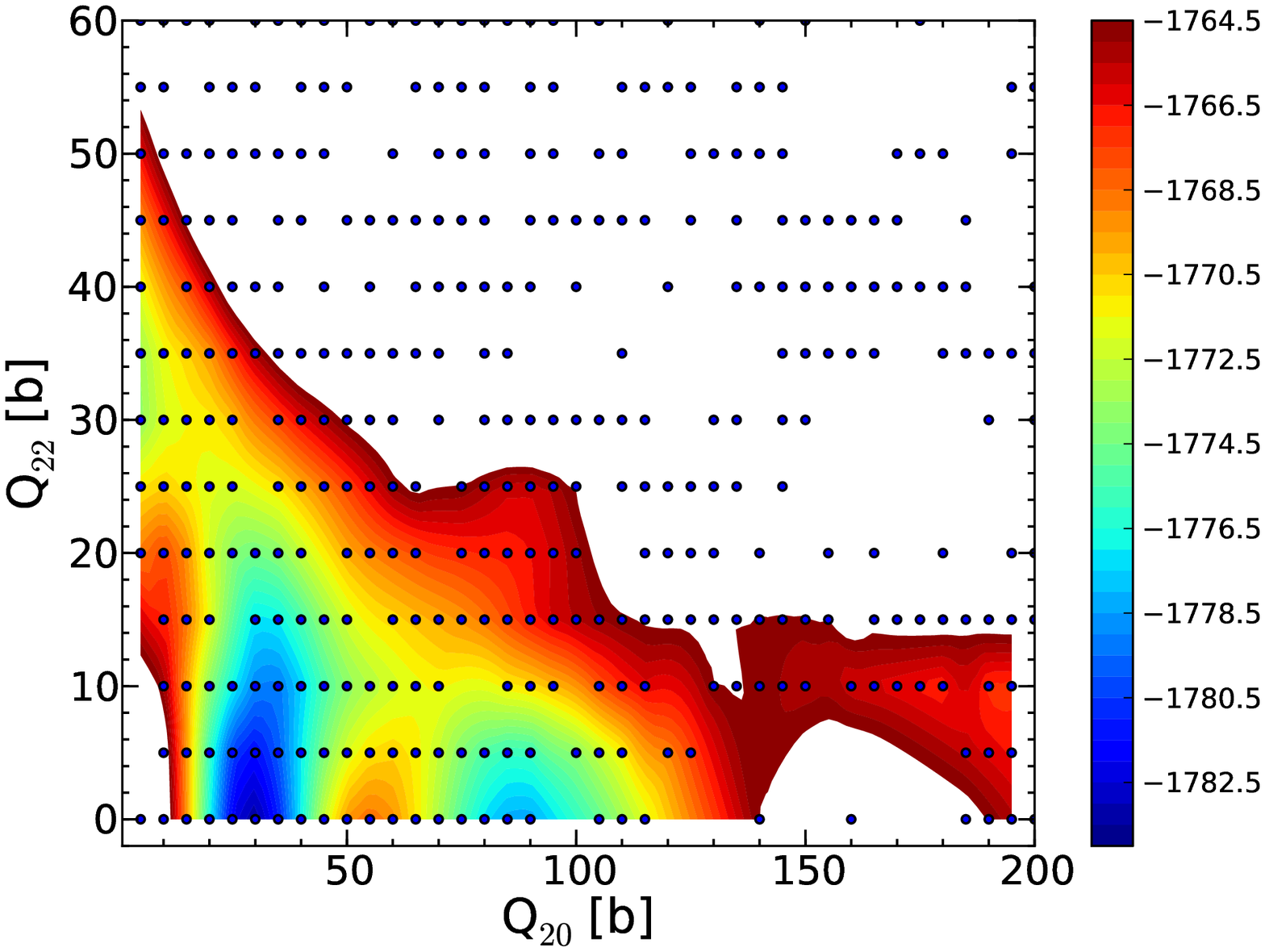}
  \caption{(color online) Small-scale illustration of convergence failure. 
           The figure shows a 2D cross-section of the potential energy 
           landscape in $^{240}$Pu. Black points are calculated points; 
           missing points did not converge.}
  \label{fig:mesh}
\end{center}
\end{minipage}
\end{center}
\end{figure}

{\sc Iterative Algorithms - } The HFB equations are usually solved by 
iterations: starting at iteration 0 with an initial guess for, for example, the 
density matrix $\rho_{\text{in}}^{(0)}$ and pairing tensor 
$\kappa_{\text{in}}^{(0)}$, one constructs the HFB matrix $M^{(0)}$; 
diagonalizing it gives the eigenvectors at iteration 0, which are used 
to compute an updated version of the density matrix $\rho_{\text{out}}^{(0)}$ 
and pairing tensor $\kappa_{\text{out}}^{(0)}$; using these updates as input 
to iteration 1, one constructs the new HFB matrix $M^{(1)}$, diagonalizes it, 
and so forth, until convergence is met. Formally, this can be written 
as
\begin{equation}
\gras{V}_{\text{out}}^{(m)} = \gras{I}(\gras{V}_{\text{in}}^{(m)}).
\end{equation}
The solution to the HFB equation satisfies: $\gras{V} = \gras{I}(\gras{V})$, 
or equivalently $\gras{F}(\gras{V}) = \gras{V} - \gras{I}(\gras{V}) = 0$. 
This is a form of the fixed-point problem. Most DFT solvers iterate 
$\gras{V}$ either with a standard linear mixing, 
\begin{equation}
\gras{V}_{\text{in}}^{(m+1)} = \alpha\gras{V}_{\text{out}}^{(m)} 
+ (1 - \alpha) \gras{V}_{\text{in}}^{(m)},
\end{equation}
or a more elaborate mixing like the modified Broyden mixing \cite{[Bar08]}.
The final number of iterations needed to reach convergence is extremely 
dependent on the type of calculation: ground-state properties of a spherical 
nucleus may take as little as 30 iterations, while the scission configuration 
in $^{240}$Pu may take as much as 5,000 iterations \cite{[You09]}. In 
addition, the iterative method often fails to converge, especially with 
large ``exotic'' constraints. Since the time of calculation is ultimately 
linearly proportional to the number of iterations, controlling the latter 
and ensuring a high convergence rate is critical for DFT applications.
		

		

\begin{figure}[h]
\center
  \includegraphics[width=0.85\linewidth]{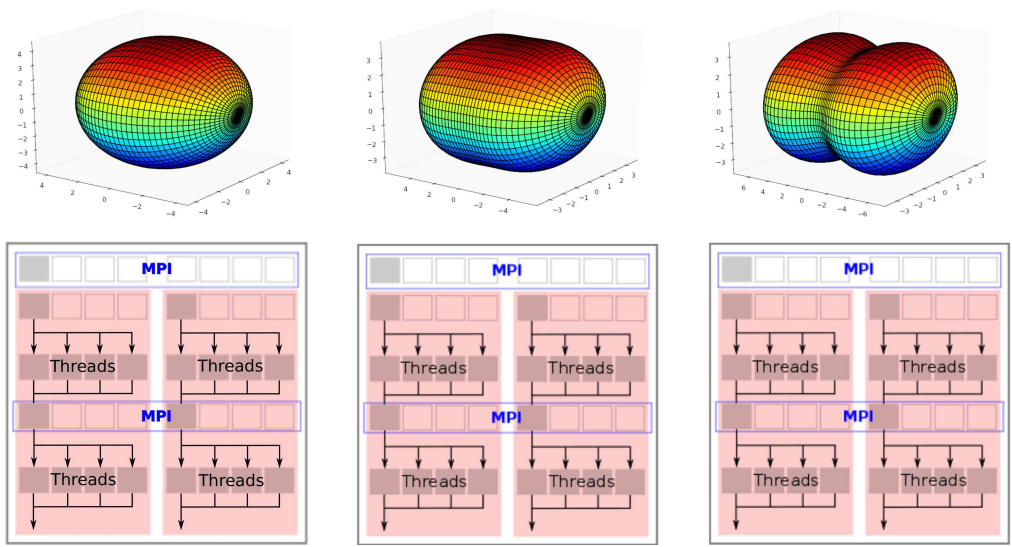}
  \caption{(color online) Hybrid MPI/OpenMP programming model for large-scale 
          DFT applications: the process grid is decomposed in MPI 
          communicators made of a few nodes (1 - 16), handling a given HFB 
          calculation, with OpenMP threading within nodes.}
  \label{fig:scheme}
\end{figure}

\subsection{Large-Scale Applications}

In itself, one HFB calculation is almost always manageable on a standard 
computer. However, realistic applications always require the computation 
of a very large number of different configurations. The static description 
of nuclear fission is a good example: at least four deformation degrees of 
freedom are necessary---elongation, triaxiality, mirror symmetry, and neck size---just 
for calculations at zero temperature and zero angular momentum. A 
typical estimate for the number of points for each degree of freedom is 
$500 \times 40 \times 20 \times 20 = 8. 10^{6}$ points. At each point, one 
can estimate the error due to the truncation of the basis by repeating the 
same calculation with several different combinations of basis parameters: 
assuming 10 points per basis parameter, this adds a factor of 1,000. The typical 
size of the problem is therefore on the order of $10^{9}--10^{12}$ independent 
calculations, each taking on the order of a few days on a single core 
for high-precision results. This estimate applies to a single nucleus only.

Modern DFT solvers have therefore adopted a hybrid MPI/OpenMP programming 
model, illustrated in Fig. \ref{fig:scheme}. Since a large number of 
HFB calculations is needed for any realistic problem, the process grid is 
decomposed into many small MPI communicators, each in charge of handling one 
HFB task, and possibly spanning multiple nodes. In order to accelerate dense linear 
algebra operations, OpenMP threading is used within a node. In many 
applications, only one MPI task per node is devoted to an HFB calculation. 
With existing solvers, the number of files needed to dump the output of the 
calculation grows as the number of HFB configurations handled: navigating 
this large mass of data and extracting the most relevant information 
pertaining to the problem at hand can be tricky. Some effort has therefore 
been put into the development of interfaces with advanced data mining 
software \cite{[Mas]}.


\section{Recent Achievements}

Collaborative efforts such as the SciDAC 2 UNEDF project have enabled 
ground-breaking optimizations of nuclear DFT solvers \cite{[Sch11]}, which 
in turn have led to important discoveries in several areas of the physics of 
heavy nuclei. We highlight below two examples of recent work.

\begin{figure}[h]
\begin{center}
\begin{minipage}[t]{0.48\textwidth}
\begin{center}
  \includegraphics[width=0.8\linewidth,angle=-90]{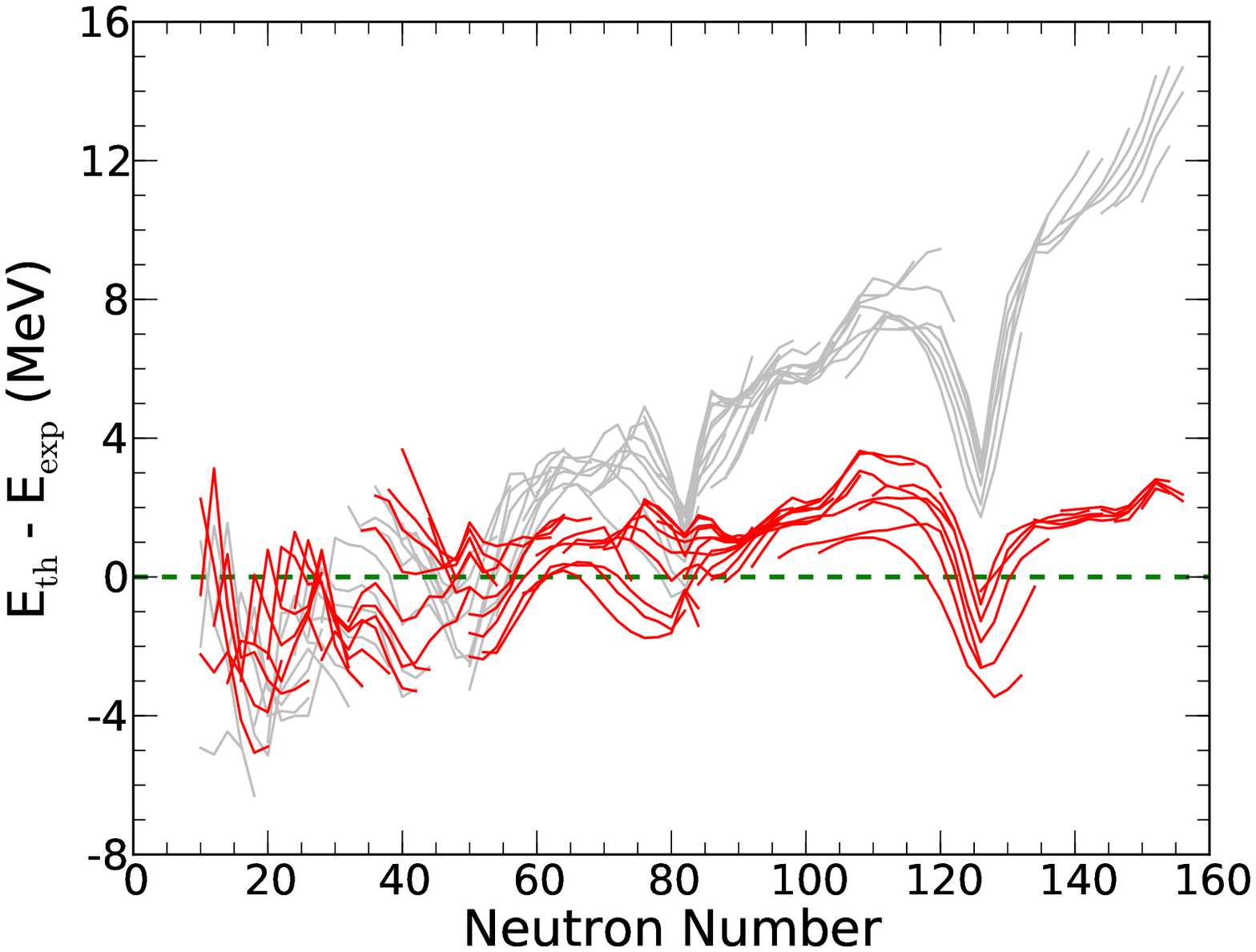}
  \caption{(color online) Differences between theoretical and experimental 
           atomic masses for SLy4 (gray) and UNEDF1 (red) parametrization. 
           Each line corresponds to an isotopic sequence. Adapted from 
           \cite{[Kor10]}.}
  \label{fig:mass}
\end{center}
\end{minipage}\hspace{0.03\textwidth}%
\begin{minipage}[t]{0.48\textwidth}
\begin{center}
  \includegraphics[width=0.8\linewidth,angle=-90]{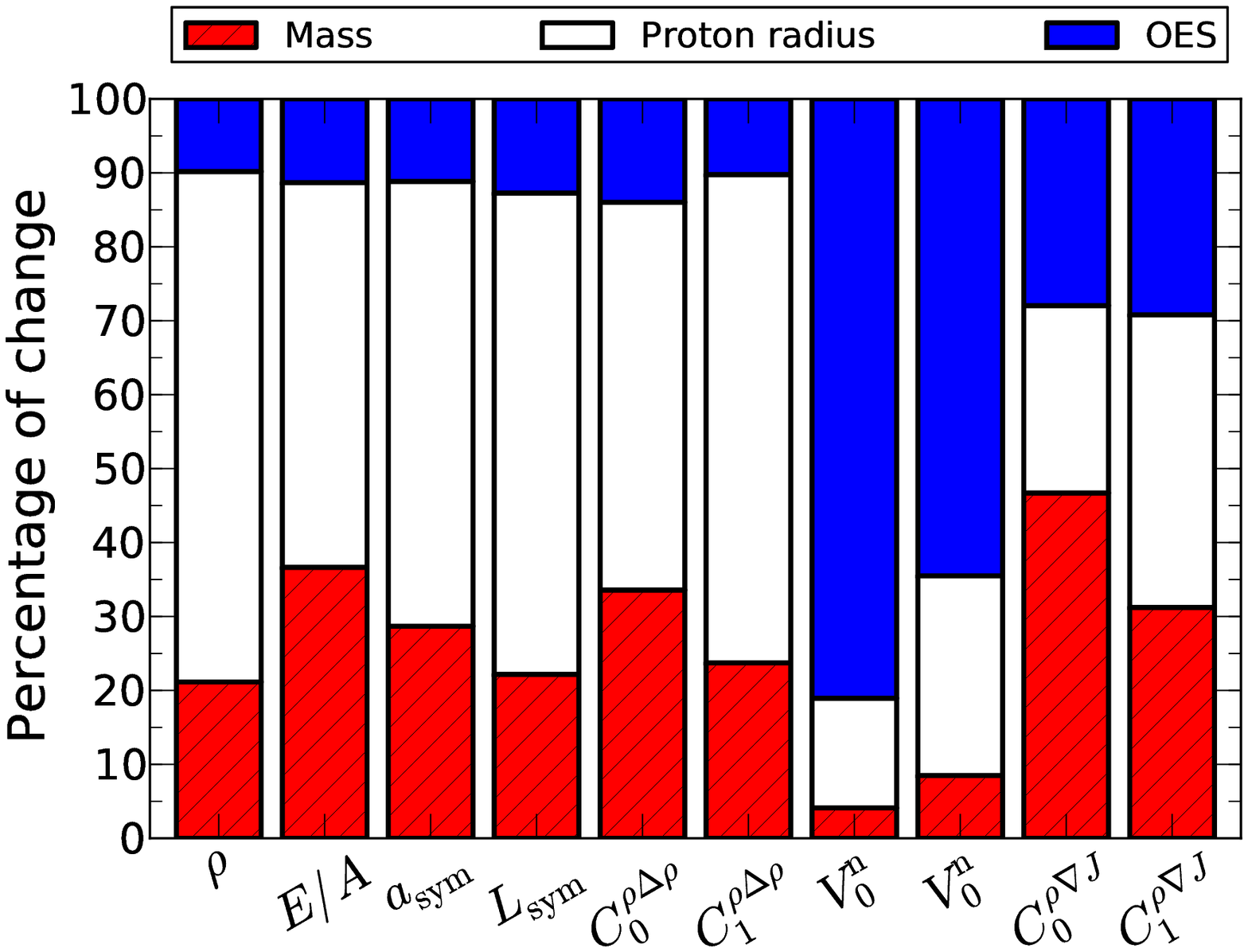}
  \caption{(color online) Sensitivity of the parameters of the Skyrme 
           functional to various types of experimental data included 
           in the optimization. Adapted from \cite{[Kor10]}. }
  \label{fig:correlations}
\end{center}
\end{minipage}
\end{center}
\end{figure}

{\sc Optimization of Energy Functionals - } The only input to nuclear DFT 
is the dozen or so low-energy constants characterizing the energy functionals. 
These parameters need to be  carefully adjusted to experimental data. 
In the past, this procedure was usually carried out for specific systems 
such as infinite nuclear matter or doubly magic spherical nuclei, essentially 
because calculations are fast for those cases. However, most realistic nuclei 
are significantly different from such idealized systems, and it has been 
realized that many energy functionals suffer from systematic biases. 
The availability of heavily optimized DFT solvers together with leadership-class computers has allowed parameter optimization to be performed in realistic 
nuclei, that is, deformed nuclei with pairing. Moreover, statistical methods 
can now be applied at the solution to investigate the sensitivity of the 
solution to the experimental data, as well as built-in correlations between 
the parameters. Such modern methods have shed new light on the validity 
of current functionals and are now being applied to new generations of 
functionals \cite{[Kor10]}.

\begin{figure}[h]
\begin{center}
\begin{minipage}[t]{0.48\textwidth}
\begin{center}
  \includegraphics[width=0.8\linewidth]{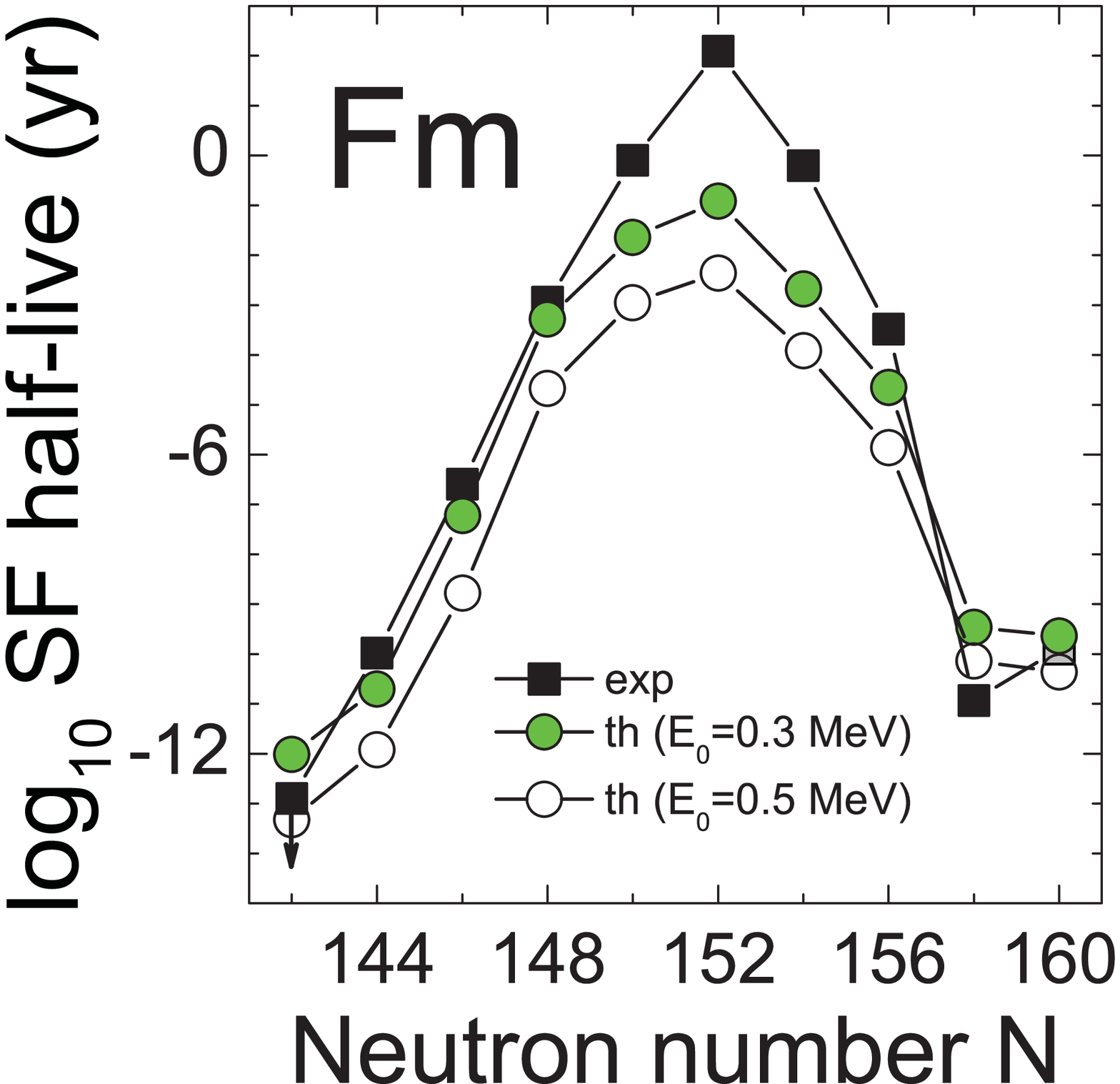}
  \caption{(color online) Theoretical and experimental fission half-lives 
           of even-even Fermium isotopes.}
  \label{fig:lifetimes}
\end{center}
\end{minipage}\hspace{0.03\textwidth}%
\begin{minipage}[t]{0.48\textwidth}
\begin{center}
  \includegraphics[width=\linewidth]{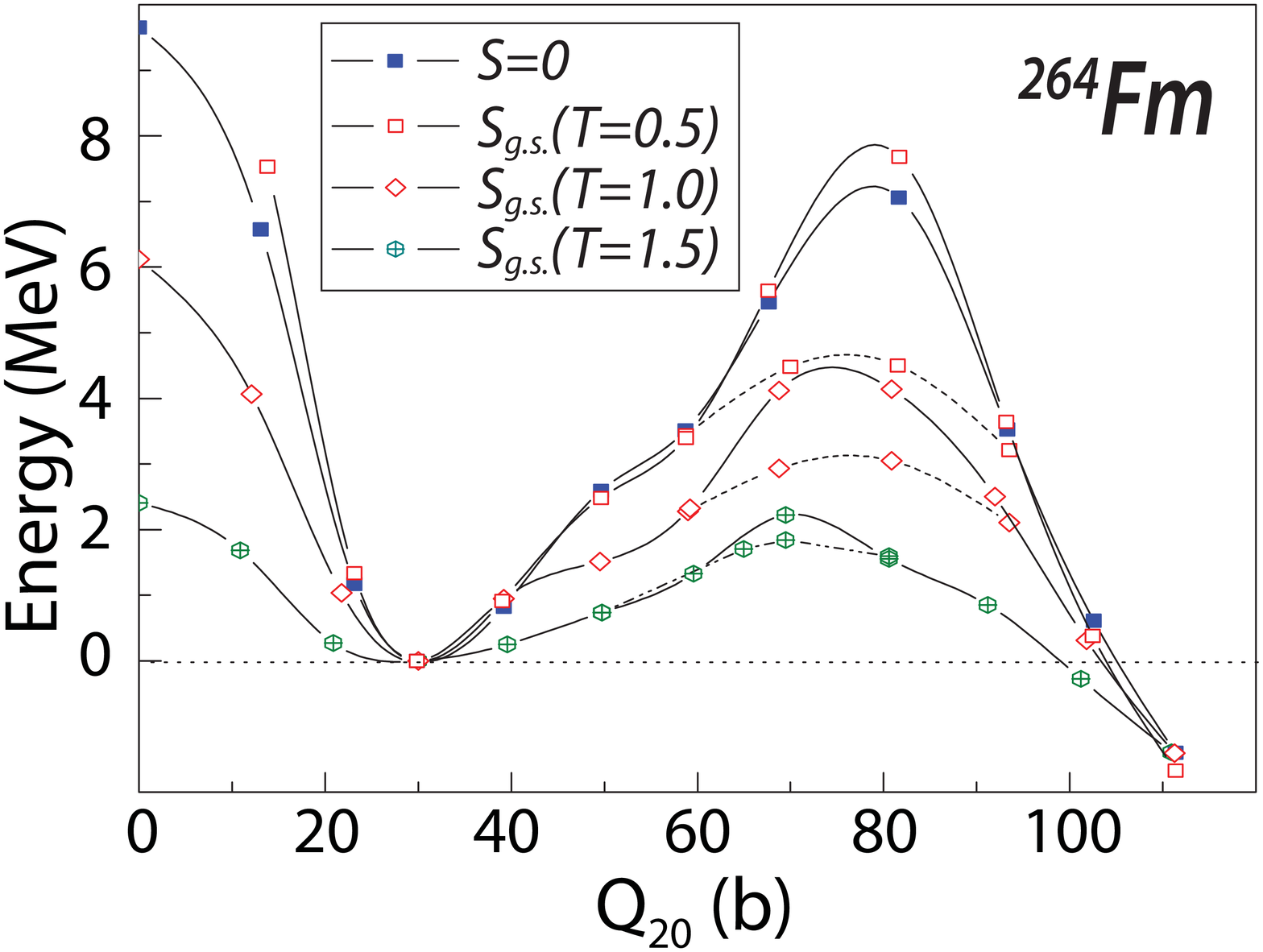}
  \caption{(color online) Symmetric isentropic fission pathways of 
           $^{264}$Fm (energy normalized to the ground-state) including 
           triaxial degrees of freedom (dashed lines).}
  \label{fig:temperature}
\end{center}
\end{minipage}
\end{center}
\end{figure}

{\sc Description of the Fission Process - } The successful description of 
the fission process in the framework of DFT is a poster-child example of a 
large-scale computational problem involving nuclear DFT that could have  
tremendous applications for society. Most of the recent progress in the 
field has come from the computational side. In particular, the first 
systematic self-consistent survey of fission pathways with several shape 
degrees of freedom has been carried out in the region of the heaviest 
elements. Calculated lifetimes are in reasonable agreement with experiment; 
see Fig. \ref{fig:lifetimes} \cite{[Sta09]}. In parallel, preliminary studies 
of compound nucleus fission have shed new light on the probability of 
formation for superheavy elements in fusion reactions; see Fig. \ref{fig:temperature} \cite{[Pei09],[She09]}.
	
\section{Moving Forward}

{\sc Computing Excited States - } Most of the examples presented in this 
article correspond to the ground-state properties of nuclei. A significant 
challenge to DFT is its ability to also describe excited states. In the 
version of DFT that derives from a two-body effective interaction, 
indications are that a three-body force will have to be explicitly included. 
Doing so would enable to apply a well-established set of techniques such as 
projection and the generator coordinate method that can provide excited 
spectra. However, these methods will require yet another leap in the number 
of HFB points to be computed: for example, tensor contractions with a three-body 
force will increase from 12 to 18 the number of nested loops needed to 
compute the mean-field $\Gamma_{ac}$ in Cartesian coordinates.
		
\begin{figure}[h]
\begin{center}
\begin{minipage}[t]{0.46\textwidth}
\begin{center}
  \includegraphics[width=0.8\linewidth,angle=-90]{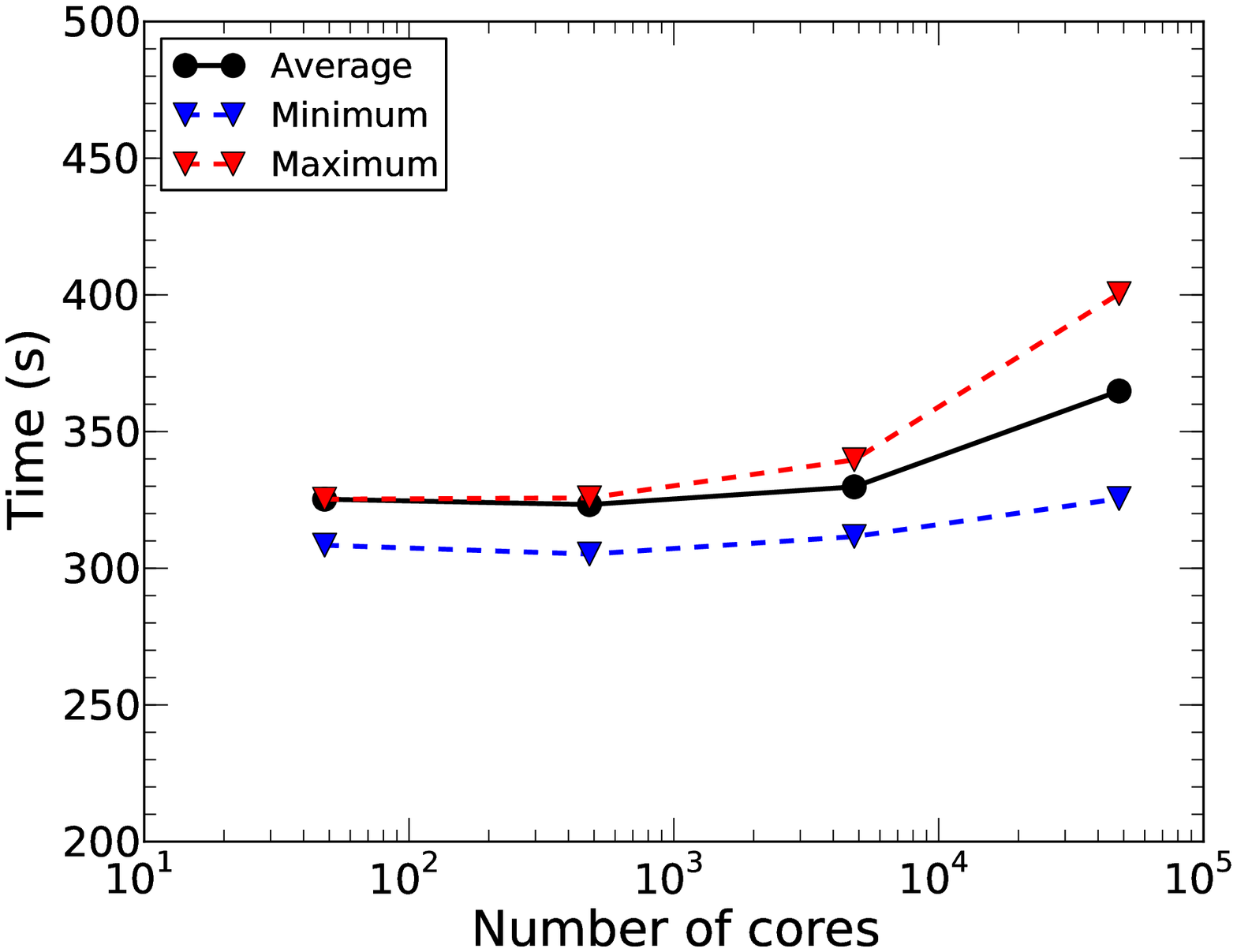}
  \caption{(color online) Scaling of the DFT solver HFODD for identical 
           calculations run in parallel. Most of the load-imbalance and 
           imperfect scaling comes from I/O. From \cite{[Sch11]}.}
  \label{fig:scaling}
\end{center}
\end{minipage}\hspace{0.03\textwidth}%
\begin{minipage}[t]{0.46\textwidth}
\begin{center}
  \includegraphics[width=0.8\linewidth,angle=-90]{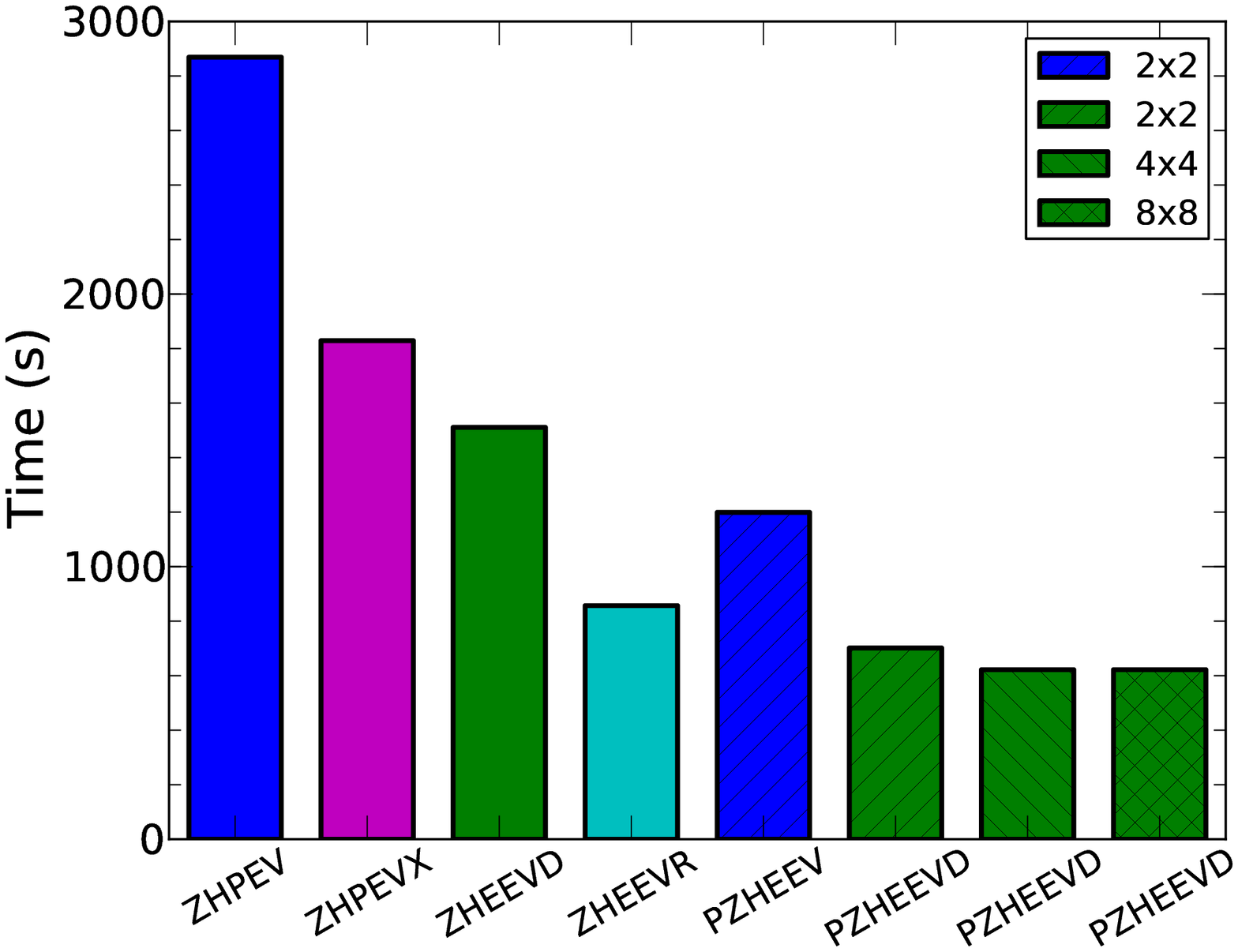}
  \caption{(color online) Time of calculation for six HFB iterations depending 
           on the ScaLAPACK diagonalization routine and the process grid 
           (matrices of rank 2760).}
  \label{fig:scalapack}
\end{center}
\end{minipage}
\end{center}
\end{figure}

{\sc Advanced Data Management - } Currently, all DFT solvers have a rather 
simple I/O system, which essentially relies on native Fortran or C/C++ 
routines for disc access. Files are written on a per core basis: in large-scale applications, the number of files becomes huge and its management 
rather complex, and the scalability may degrade quickly as one hits 
the limits of the operating system; see Fig. \ref{fig:scaling}. It seems 
therefore necessary to invest into more efficient I/O systems, possibly 
interfaced with professional database management and data-mining software. 
Indeed, a specific feature of DFT is that it produces a lot of data points 
that need to be analyzed in many different ways. 

{\sc Real-time Simulation Steering - } Current large-scale simulations are 
intrinsically static: given a set of input data shared among 
processes, each group of cores performs its task until completion. However, 
entire regions of potential energy surfaces irrelevant for physics 
applications cannot be detected until postanalysis is performed; 
calculations that failed could be converged with, for example, slightly different 
mixing parameters; model space dependence could be efficiently studied by 
optimizaton over the basis parameters, rather than meshing the parameter 
space. All these observations point to the need of dynamically steering 
the simulation based on a set of preliminary results. However, such a 
program can be viable only if the time of one HFB calculation can be reduced 
to at least less than 1 hour.

{\sc Parallel Dense Linear Algebra Libraries - } The reduction of the 
typical computation time below the 1-hour barrier is not possible without 
the development of highly optimized parallel dense linear algebra libraries. 
The current ScaLAPACK library provides a good starting point, but many 
routines do not reach the same level of performance as the original 
LAPACK versions. Currently, the gain is probably not sufficient for 
many practical application; see Fig. \ref{fig:scalapack}.


\section{Conclusions}

Nuclear DFT is the only theoretical framework that can be applied to all 
nuclei from the lightest to the heaviest, including stellar environments. 
While relatively mature, the theory has only recently started to benefit 
from the availability of leadership-class computers and advances in code 
development. Significant progress has been achieved, in particular in terms 
of parameter optimization and specific applications such as fission or 
large-scale surveys \cite{[Sch10]}. It is reasonable to anticipate that 
numerical uncertainties due to the truncation of the model space (HO basis) 
and the collective space (number of deformations) could be virtually 
eliminated in the near future, which would then open the door to 
high-precision nuclear simulations of importance to science and society. 
Such a promise, though, can be delivered only by the joint effort of both 
nuclear scientists and computer scientists. On the computational side, some 
of the key aspects are the parallelization of dense linear algebra operations, 
the development of real-time simulation steering tools, and the implementation of 
scalable I/O models and data management tools. 


\section*{Acknowledgments}
This work was supported by the Office of Nuclear Physics, U.S. Department of 
Energy under Contract Nos. DE-FC02-09ER41583 (UNEDF SciDAC Collaboration), 
DE-FG02-96ER40963 and DE-FG02-07ER41529 (University of Tennessee), 
DE-FG0587ER40361 (Joint Institute for Heavy Ion Research) and DE-AC02-06CH11357 
(Argonne National Laboratory), and by the Polish Ministry of Science and Higher Education Contract NN202231137. It was partly performed under the auspices of 
the US Department of Energy by the Lawrence Livermore National Laboratory under 
Contract DE-AC52-07NA27344. Funding was also provided by the United States 
Department of Energy Office of Science, Nuclear Physics Program pursuant to 
Contract DE-AC52-07NA27344 Clause B-9999, Clause H-9999 and the American 
Recovery and Reinvestment Act, Pub. L. 111-5. Computational resources were 
provided through an INCITE award ``Computational Nuclear Structure'' by the 
National Center for Computational Sciences (NCCS) and National Institute for 
Computational Sciences (NICS) at Oak Ridge National Laboratory, and through an 
award by the Laboratory Computing Resource Center (LCRC) at Argonne National 
Laboratory.



\section*{References}

\begin{thebibliography}{10}

\bibitem{[Ben03]}
{M. Bender, P.-H. Heenen, and P.-G. Reinhard, Rev. Mod. Phys. {\bf 75}, 121
  (2003).}

\bibitem{[Dob07d]}
{J. Dobaczewski, M.V. Stoitsov, W. Nazarewicz, and P.-G. Reinhard, C {\bf 76},
  054315 (2007).}

\bibitem{[Dug10]}
{T. Duguet and J. Sadoudi, J. Phys. G: Nucl. Part. Phys. {\bf 37} 064009
  (2010).}

\bibitem{[Ben05]}
{K. Bennaceur and J. Dobaczewski, Comput. Phys. Commun. {\bf 168}, 96 (2005)}.

\bibitem{[Pei08]}
{J. C. Pei, M. V. Stoitsov, G. I. Fann, W. Nazarewicz, N. Schunck, and F. R.
  Xu, Phys. Rev. C {\bf 78}, 064306 (2008).}

\bibitem{[Nik10]}
{N. Nikolov, N. Schunck, W. Nazarewicz, M. Bender, and J. Pei, Phys. Rev. C
  {\bf 83}, 034305 (2011).}

\bibitem{[Fan09]}
{G.I. Fann, J. Pei, R.J. Harrison, J. Jia, J. Hill, M. Ou, W. Nazarewicz, W. A.
  Shelton, and N. Schunck, J. Phys, Conference Series; {\bf 180} 012080 (2009)
  Proc. SciDAC 2009 Conference, San Diego, CA.}

\bibitem{[Dob09d]}
{J. Dobaczewski, W. Satu{\l}a, B.G. Carlsson, J. Engel, P. Olbratowski, P.
  Powa{\l}owski, M. Sadziak, J. Sarich, N. Schunck, A. Staszczak, M.V.
  Stoitsov, M. Zalewski, and H. Zdu\'nczuk, Comput. Phys. Commun. {\bf 180},
  2361 (2009)}.

\bibitem{[Bar08]}
{A. Baran, A. Bulgac, M. McNeil Forbes, G. Hagen, W. Nazarewicz, N. Schunck,
  and M.V. Stoitsov, Phys. Rev. C {\bf 78}, 014318 (2008).}

\bibitem{[You09]}
{W. Younes and D. Gogny, Phys. Rev. C {\bf 80}, 054313 (2009).}

\bibitem{[Mas]}
{http:$\backslash\backslash$massexplorer.org.}

\bibitem{[Sch11]}
{N. Schunck, J. Dobaczewski, J. McDonnell, W. Satu{\l}a, J.A. Sheikh, A.
  Staszczak, M. Stoitsov, P. Toivanen, arXiv:1103.1851 (2011).}

\bibitem{[Kor10]}
{M. Kortelainen, T. Lesinski, J. Mor\'e, W. Nazarewicz, J. Sarich, N. Schunck,
  M.V. Stoitsov, and S. Wild, Phys. Rev. C {\bf 82}, 024313 (2010).}

\bibitem{[Sta09]}
{A. Staszczak, A. Baran, J. Dobaczewski and W. Nazarewicz, Phys. Rev. C {\bf
  80}, 014309 (2009).}

\bibitem{[Pei09]}
{J.C. Pei, W. Nazarewicz, J.A. Sheikh, A.K. Kerman, Phys. Rev. Lett. {\bf 102},
  192501 (2009).}

\bibitem{[She09]}
{J.A. Sheikh, W. Nazarewicz, J.C. Pei, Phys. Rev. C {\bf 80}, 011302 (2009).}

\bibitem{[Sch10]}
{N. Schunck, J. Dobaczewski, J. Mor\'e, J. McDonnell, W. Nazarewicz, J. Sarich,
  and M. V. Stoitsov, Phys. Rev. C {\bf 81}, 024316 (2010).}

\end{thebibliography}
\end{document}